\documentclass[11pt]{article}
\usepackage{epsfig}
\usepackage{latexsym}
\title{Quantum Corrections to the Schwarzschild and Kerr Metrics}
\author{N. E. J. Bjerrum-Bohr$^a$\\
John F. Donoghue$^b$, and Barry R. Holstein$^{b,c}$\\
$^a$ University of Copenhagen\\
The Niels Bohr Institute\\
Blegdamsvej 17, Copenhagen \O \\
DK-2100, Denmark\\
$^b$ Department of Physics\\
University of Massachusetts\\
Amherst, MA  01003\\
$^c$ Institut f\"{u}r Kernphysik\\
Forschungszentrum J\"{u}lich\\
D-52425 J\"{u}lich, Germany}
\begin{document}
\begin{titlepage}
\maketitle
\begin{abstract}
We examine the corrections to the lowest order gravitational
interactions of massive particles arising from gravitational
radiative corrections. We show how the masslessness of the
graviton and the gravitational self interactions imply the
presence of nonanalytic pieces $\sim\sqrt{-q^2},\sim q^2\ln-q^2$,
etc. in the form factors of the energy-momentum tensor and that
these correspond to long range modifications of the metric tensor
$g_{\mu\nu}$ of the form $G^2m^2/r^2,G^2m\hbar/r^3$, etc.  The
former coincide with well known solutions from classical general
relativity, while the latter represent new quantum mechanical
effects, whose strength and form is necessitated by the low energy
quantum nature of the general relativity.  We use these results to
define a running gravitational charge.
\end{abstract}
\end{titlepage}

\section{Introduction}
In this paper we will discuss the long distance classical and quantum
corrections to the Schwarszchild and Kerr
metrics using the techniques of effective field theory. We will show how
the nonanalytic radiative corrections to
the energy-momentum tensor can be used to obtain the classical nonlinear
terms in these metrics at long distance,
and calculate the analogous nonanalytic quantum corrections. For the
Schwarzschild metric we consider the case of
a massive scalar particle. Here we clear up some numerical disagreements
in related calculations that have emerged
in the literature. We then present the Kerr results, using a massive
fermion as a source, and show that the
spin-independent quantum corrections are the same as those of the scalar
particle. We also elucidate various
theoretical issues and compare with other results in the literature.

Effective field theory is ideally suited for discussing the quantum
effects of general relativity at scales well
below the Planck mass\cite{don,weinberg}. While it is expected that the
degrees of freedom and the interactions of
gravity will be modified beyond the Planck scale, at low energies these
ingredients are accurately described by
general relativity. Effective field theory separates the known quantum
effects of the low energy particles from
the unknown physics at high energy. The latter effects are represented
by the most general series of effective
lagrangians consistent with the symmetry of general relativity. However
the propagation of the low energy
particles yields identifiable quantum effects that can be isolated by
the techniques of effective field theory.

The present study builds on two sets of recent work. One of these is the
use of effective field theory to study
quantum corrections to the gravitational potential\cite{don,
don1,akh,bohr,hamber,muzinich,nejbb}. While the basic
principles of these studies are the same, there are some differences
and/or disagreements. Since there is not a
universal definition of the meaning of a potential, different authors
use different definitions of potential in
terms of Feynman diagrams, and hence obtain different answers. Even in a
case where the same definition is used,
different results have been obtained. We will provide some clarification
of these disagreements. A related paper
\cite{nejbb} provides a full and detailed calculation
of the scattering potential of scalar
particles. In the present paper we note that a subset of diagrams is
more readily interpreted as a change in the
metric, and we calculate these effects.

The other precedent for the present paper is the calculation of
the the leading quantum corrections to the Reissner-Nordstrom and
Kerr-Newman metrics using effective field
theory\cite{pa1}\footnote{These corrections have also been
considered from the point of view of S-matrix theory in ref.
\cite{nbj}}. These metrics involve charged particles, so that the
quantum corrections involved photon loops, not graviton loops.
However, this provided a particularly clear laboratory for the
study of metric corrections. Interestingly, we saw that the
\emph{classical} nonlinearities in the metric can be calculated
straightforwardly using Feynman diagram techniques. At the same
time we saw that there was a clear identification of certain
nonanalytic terms with long-distance quantum effects in the
metric. We use the insights of that study to investigate the
present problems, which involve graviton loops. In the present
case the interpretation is not as clear, although the calculations
are well defined.

We will be using harmonic gauge throughout this paper. In this gauge,
the Schwarzschild metric has the
form\cite{wein,sch},
\begin{eqnarray}
g_{00}&=&\left(1-{Gm\over r}\over 1+{Gm\over r}\right)=1-2{Gm\over r}
+2{G^2m^2\over r^2}+\ldots\nonumber\\
g_{0i}&=&0\nonumber\\
g_{ij}&=&-\delta_{ij}(1+{Gm\over r})^2-{G^2m^2\over r^2}\left(1+{Gm\over
r}
\over 1-{Gm\over r}\right){r_ir_j\over r^2}\nonumber\\
&=&-\delta_{ij}\left(1+2{Gm\over r}+{G^2m^2\over r^2}\right)
-{r_ir_j\over r^2}{G^2m^2\over
r^2}+\ldots\label{eq:sch}
\end{eqnarray}
The Kerr metric\cite{ker} refers to a particle with spin and, keeping
only terms up to first order in the angular
momentum, has the harmonic gauge form
\begin{eqnarray}
g_{00}&=&\left(1-{Gm\over r}\over 1+{Gm\over
r}\right)+\ldots=1-2{Gm\over r}
+2{G^2m^2\over r^2}+\ldots\nonumber\\
g_{0i}&=&{2G\over r^2(r+mG)}(\vec{S}\times\vec{r})_i+\ldots=\left(
{2G\over r^3}-{2G^2m\over
r^4}\right)(\vec{S}\times\vec{r})_i+\ldots\nonumber\\
g_{ij}&=&-\delta_{ij}(1+{Gm\over r})^2-{G^2m^2\over r^2}\left(1+{Gm\over
r}
\over 1-{Gm\over r}\right){r_ir_j\over r^2}+\ldots\nonumber\\
&=&-\delta_{ij}\left(1+2{Gm\over r}+{G^2m^2\over r^2}\right)
-{r_ir_j\over r^2}{G^2m^2\over
r^2}+\ldots\label{eq:ker}
\end{eqnarray}
We will show that using a particular set of Feynman diagrams we
reproduce the former with the addition of a long
distance quantum correction
\begin{eqnarray}
g_{00}&=&1-2{Gm\over r}
+2{G^2m^2\over r^2}+{62G^2 m \hbar\over 15\pi r^3} +\ldots\nonumber\\
g_{0i}&=& 0\nonumber\\
g_{ij} &=&-\delta_{ij}\left(1+2{Gm\over r}+{G^2m^2\over r^2}+{14G^2 m
\hbar\over 15\pi r^3}\right) -{r_ir_j\over
r^2}\left({G^2m^2\over r^2}+{76G^2 m \hbar\over 15\pi
r^3}\right)+\ldots\label{eq:sh2}
\end{eqnarray}
For the Kerr metric,
\begin{eqnarray}
g_{00}&=&1-2{Gm\over r}
+2{G^2m^2\over r^2}+{62G^2 m \hbar\over 15\pi r^3}+\ldots\nonumber\\
g_{0i}&=&\left( {2G\over r^3}-{2G^2m\over r^4}+{36G^2\hbar\over 15\pi
r^5}\right)(\vec{S}\times\vec{r})_i+\ldots\nonumber\\
g_{ij}&=&-\delta_{ij}\left(1+2{Gm\over r}+{G^2m^2\over r^2}+{14G^2 m
\hbar\over 15\pi r^3}\right) -{r_ir_j\over
r^2}\left({G^2m^2\over r^2}+{76G^2 m \hbar\over 15\pi
r^3}\right)+\ldots\label{eq:ker2}
\end{eqnarray}
It is of course required that the classical spin-independent terms must
be the same for a scalar particle and a
fermion. We know of no firm requirement for the spin-independence of the
quantum corrections to $g_{00}$ and
$g_{ij}$, but from our calculation they are seen to be identical.

\section{Review}
The metric is derived from the energy-momentum tensor of a source, using
Einstein's equation as the equation of
motion. At lowest order in the fields, the source particle is just a
point particle in coordinate space. However,
both classical fields and their quantum fluctuations modify the
energy-momentum of a particle at long distance.
These modifications can be found by the consideration of the radiative
corrections to the energy-momentum tensor.
When these are translated into a metric, they yield the classical
nonlinearities and quantum modifications of the
metric.

Let us review what was found in Ref\cite{pa1} for the conceptually
simpler case of charged particles, as our
calculation here will follow the same procedure. In that case the field
around the particles was the
electromagnetic field and the gravitational interaction was treated
purely classically. The masslessness of the
photon implies that there are long range fields around a charged
particle and these carry energy and momentum. At
the same time, in a Feynman diagram calculation of the renormalization
of the energy momentum tensor of the
charged particle, the masslessness of the photon leads to nonanalytic
terms in the formfactors having the
structure $\sim\sqrt{-q^2},\sim q^2\ln-q^2$, where $q$ is the momentum
transfer, as well as analytic terms of
order $q^2,q^4...$. It was shown in detail how the $\sim\sqrt{-q^2}$
terms account for the classical
energy-momentum of the electromagnetic field and how they exactly
reproduce the classical nonlinearities in the
metric that are present in the Reissner Nordstrom and Kerr Newman
metrics. Nonanalytic terms of the form
$q^2\ln-q^2$ also appear, and when they are included in the equations of
motion they produce further corrections
in these metrics. Explicit examination shows that these latter are
linear in $\hbar$ - i.e.they are quantum
effects. Finally the analytic terms produce only delta functions (or
derivatives of delta functions) in the
metric, such that they vanish at long distance. Thus we saw that the
long distance modifications of the metrics
are obtained from the nonanlaytic terms in the formfactors of the energy
momentum tensor.

The same effects are present in the purely gravitational case. If one
expands the energy and momentum of the
particle in powers of G, the lowest order result is that of a point
particle. However, there is energy and
momentum also carried by the gravitational field around the particle and
this can be calculated via the one loop
Feynman diagrams. Because the graviton is massless, there will also be
nonanalytic terms of the forms
$\sim\sqrt{-q^2},\sim q^2\ln-q^2$ in the formfactors of the energy
momentum tensor. Again these will produce long
range modifications of the metric. If we include the relevant
dimensionful couplings, this will have the
schematic form
   \begin{eqnarray}
   {\rm metric} &\sim& Gm\int {d^3q \over (2\pi)^3 }
 e^{i\vec{q}\cdot\vec{r}} {1 \over \vec{q}^2}
   \left[ 1- a G
   {\vec{q}^2 }\sqrt{m^2 \over \vec{q}^2} - b G
   {\vec{q}^2 } \log(\vec{q}^2) - c G {\vec{q}^2 }+\ldots
   \right] \nonumber \\
   &\sim& Gm \left[ {1\over r} +{a G m\over  r^2} +{ b G  \hbar\over
   r^3}
   +{c G} \delta^3(x) +\ldots \right]
   \end{eqnarray}
Here $a,b,c$ are dimensionless numbers and further numerical factors of
order unity will be inserted later. We
will examine the $\sqrt{-q^2}$ terms in detail and show how they
correctly reproduce all the features of the
classical metric. For the $q^2\ln-q^2$ terms we have included the factor
of $\hbar$ that follows from dimensional
analysis. The analytic correction to the energy momentum tensor yields
the delta function term that is not
relevant for the long distance behavior.

The nonanalytic terms come from the low energy propagation of gravitons,
using the couplings of general
relativity. Because these features are independent of the high energy
behavior of gravity, they are unambiguous
predictions of low energy general relativity. There is also no influence
of other possible terms in the
gravitational lagrangian, such as $R^2$ or related corrections in the
matter lagrangian. These yield only analytic
corrections to the formfactor and hence do not provide long distance
modifications of the metric. These are
behaviors that are well known in the effective field theory of
gravity\cite{don}.

\section{Lowest order}

Let us first consider the theory without loop corrections.  The metric
tensor is expanded as
\begin{equation}
g_{\mu\nu}\equiv\eta_{\mu\nu}+h^{(1)}_{\mu\nu}+\ldots
\end{equation}
where $\eta_{\mu\nu}=(1,-1,-1,-1)_{\rm diag}$ is the usual Minkowski
metric and the superscript refers to the number of powers of the
gravitational coupling which appear. The dynamical relation which connects
$h_{\mu\nu}^{(1)}$ and the energy momentum tensor $T_{\mu\nu}$ is the
Einstein equation, whose linearized form in
harmonic gauge---$g^{\mu\nu}\Gamma^\lambda_{\mu\nu}=0$---is
\begin{equation}
\Box h_{\mu\nu}^{(1)}=-16\pi G(T_{\mu\nu}(x)-{1\over
2}\eta_{\mu\nu}T(x))\label{eq:ein}
\end{equation}
where $T=\eta^{\mu\nu}T_{\mu\nu}$ represents the trace. The metric for a
nearly static source is then recovered
via the Green function in either coordinate or momentum space
\begin{eqnarray}
h_{\mu\nu}(x) &=& -16\pi G \int d^3y D(x-y) (T_{\mu\nu}(y)-{1\over
2}\eta_{\mu\nu}
   T(y))  \nonumber\\
&=& -16\pi G \int {d^3q \over (2\pi)^3} e^{i\vec{q}\cdot\vec{r}}{1\over
\vec{q}^2}(T_{\mu\nu}(q)-{1\over
2}\eta_{\mu\nu}
   T(q))
\end{eqnarray}

For a quantum mechanical system, $T_{\mu\nu}$ is given in terms of the
transition density
$$<p_2|T_{\mu\nu}(x)|p_1>$$
and the conservation condition $\partial^\mu T_{\mu\nu}=0$ together with
the requirement that $T_{\mu\nu}$
transform as a second rank Lorentz tensor demands the general (scalar
field) form\footnote{Here we use the
conventional normalization for the scalar field
\begin{equation}
<p_2|p_1>=2E_1(2\pi)^3\delta^3(\vec{p}_2-\vec{p}_1)
\end{equation}}
\begin{equation}
<p_2|T_{\mu\nu}(x)|p_1>={e^{i(p_2-p_1)\cdot x}\over
\sqrt{4E_2E_1}}\left[2P_\mu P_\nu F_1(q^2)+(q_\mu
q_\nu-\eta_{\mu\nu} q^2)F_2(q^2)\right]
\end{equation}
where we have defined $P_\mu={1\over 2}(p_1+p_2)_\mu$ and
$q_\mu=(p_1-p_2)_\mu$.  Conservation of energy and
momentum requires $F_1(q^2=0)=1$ but there exists no constraint on
$F_2(q^2)$.

For the case of a point mass $m$ the lowest order form is ({\it cf.} Eq.
\ref{eq:mlg})
\begin{equation}
<p_2|T_{\mu\nu}^{(0)}(0)|p_1>={1\over \sqrt{4E_2E_1}}\left[2P_\mu P_\mu
-{1\over 2}(q_\mu
q_\nu-\eta_{\mu\nu}q^2)\right]\label{eq:emom}
\end{equation}
while in the case of the spin 1/2 system we have
\begin{eqnarray}
<p_2|T_{\mu\nu}^{(0)}(0)|p_1>&=&\bar{u}(p_2){1\over
2}\left(\gamma_\mu P_\nu
+\gamma_\nu P_\mu\right)u(p_1)\nonumber\\
&=&\left[{1\over m}P_\mu P_\nu -{i\over
4m}(\sigma_{\mu\lambda}q^\lambda
P_\nu+\sigma_{\nu\lambda}q^\lambda P_\mu)\right]u(p_1)\nonumber\\
\quad
\end{eqnarray}
where we use here the conventions of Bjorken and Drell and have
employed the Gordon identity\cite{bjd}. In either case, for a
heavy point mass located at the origin we have the lowest order
Breit frame result
\begin{equation}
<p_2|T_{\mu\nu}^{(0)}(0)|p_1>\simeq m\delta_{\mu 0}\delta_{\nu 0}
\end{equation}
The Einstein equation then has the solution
\begin{equation}
h_{\mu\nu}^{(1)}(\vec{q})=-{8\pi Gm\over \vec{q}^2}\times\left\{
\begin{array}{cc}
1& \mu=\nu=0\\ 0&\mu=0,\nu=i\\
\delta_{ij}&\mu=i,\nu=j\end{array}\right.+\ldots
\end{equation}
which, using
\begin{equation}
\int {d^3q\over (2\pi)^3}e^{i\vec{q}\cdot\vec{r}}{1\over
\vec{q}^2}={1\over 4\pi r},\quad \int {d^3q\over
(2\pi)^3} e^{i\vec{q}\cdot\vec{r}}{q_j\over \vec{q}^2}={ir_j\over 4\pi
r^3},\label{eq:r0}
\end{equation}
corresponds to the coordinate space result\footnote{Here the ellipses
represent a very short range component
associated with the q-dependent piece of $T_{\mu\nu}$.}
\begin{equation}
h_{\mu\nu}^{(1)}(\vec{r})=f(r)\times\left\{
\begin{array}{cc}1& \mu=\nu=0\\ 0&\mu=0,\nu=i\\
\delta_{ij}&\mu=i,\nu=j\end{array}\right.+\ldots\label{eq:sch1}
\end{equation}
with $$f(r)=-{2Gm\over r}$$ and reproduces the well-known leading order
piece of the Schwarzschild
solution\cite{sch}. In the case of spin 1/2 there is an additional
classical component which arises from the spin.  Using
\begin{equation}
<p_2|T_{0i}^{(0)}(0)|p_1>\simeq
\chi_2^\dagger\vec{\sigma}\chi_1\times\vec{q}
\end{equation}
we find the off-diagonal component of the metric
\begin{equation}
h_{0i}^{(1)}(\vec{q})=-8\pi i G{1\over \vec{q}^2}(\vec{S}\times\vec{q})_i
\end{equation}
which corresponds to the coodinate space result
\begin{equation}
h_{0i}^{(1)}(\vec{r})={2G\over r^3}(\vec{S}\times\vec{r})_i
\end{equation}
and agrees to this order with the Kerr metric\cite{ker}.
With this basic material in hand we now proceed to
the inclusion of loop corrections.

\section{Loop Corrections to the Energy Momentum Tensor----Spin 0}
Of course, the lowest order discussion given above is straightforward
and familiar, while the purpose of the
present paper is determine the nonanalytic corrections $\sim
\sqrt{-q^2},\,q^2\log-q^2$ to the form factors
arising from the higher order gravitational self-interaction. The
appearance of such terms was found in I to be
associated with the feature that the graviton couples to the (massless)
photon, and the same is expected to happen
in the case of gravitational self-interaction since the graviton is
itself massless.  The relevant diagrams are
shown in Figure 1 and are similar to their electromagnetic analog
considered in I, although the tensor nature of
the graviton makes the calculation {\it considerably} more tedious.
Details of the calculation are given in
Appendix A and the results are\cite{don,bohr}
\begin{eqnarray}
F_1(q^2)&=&1+{Gq^2\over \pi}(-{3\over 4}\log{-q^2\over m^2}+
{1\over 16}{\pi^2 m\over \sqrt{-q^2}})+\ldots\nonumber\\
F_2(q^2)&=&-{1\over 2}+{Gm^2\over \pi}(-2\log{-q^2\over m^2}+{7\over
8}{\pi^2m\over \sqrt{-q^2}})+\ldots
\end{eqnarray}
As found in the case of the electromagnetically corrected vertex studied
in I, we observe that the $q^2=0$ value
of the leading form factor $F_1(q^2)$ is unchanged from its lowest order
value of unity, as required by
energy-momentum conservation, while the form factor $F_2(q^2)$, which is
not protected, {\it is} modified.  Such
higher order corrections are to be expected from the feature that
gravity is nonlinear and must contain terms to
all orders in the gravitational coupling.

The momentum space form factors imply a coordinate space structure of
the energy momentum tensor which is modified
at large distance. Using the integrals listed in Appendix A, we find the
correction to the lowest order
energy-momentum tensor to be
\begin{eqnarray}
T_{00}(\vec{r})&=&\int{d^3q\over
(2\pi)^3}e^{i\vec{q}\cdot\vec{r}}\left(mF_1({q}^2)+
{\vec{q}^2\over 2m}F_2(q^2)\right)\nonumber\\
&=&\int{d^3q\over (2\pi)^3}e^{i\vec{q}\cdot\vec{r}}\left[m+\pi
Gm^2(-{1\over 16}+{7\over 16})|\vec{q}|+{Gm\over
\pi}\vec{q}^2\log{\vec{q}^2}({3\over
4}-1)\right]\nonumber\\
&=&m \delta^3(r)-{3Gm^2\over 8\pi r^4}-{3Gm\hbar \over 4\pi^2
r^5}\nonumber\\
T_{0i}(\vec{r})&=&0\nonumber\\
T_{ij}(\vec{r})&=&{1\over 2m}\int{d^3q\over
(2\pi)^3}e^{i\vec{q}\cdot\vec{r}}(q_iq_j-\delta_{ij}\vec{q}^2)F_2
(q^2)\nonumber\\
&=&\int{d^3q\over (2\pi)^3}e^{i\vec{q}\cdot\vec{r}} \left[{7\pi
Gm^2\over
16|\vec{q}|}(q_iq_j-\delta_{ij}\vec{q}^2)-
(q_iq_j-\delta_{ij}\vec{q}^2) {Gm\over \pi}\log \vec{q}^2
\right]\nonumber\\
&=& -{7Gm^2\over 4\pi r^4}\left({r_ir_j\over r^2}-{1\over
2}\delta_{ij}\right)+ {2Gm\hbar \over \pi^2 r^5}\delta_{ij}\nonumber\\
\label{eq:ta}
\end{eqnarray}
We have inserted factors of $\hbar$ where appropriate, although we
continue to use $c=1$ units.

Note that the leading correction to $T_{\mu\nu}$ is classical in nature,
since there are no factors of $\hbar$. We
can show that this effect is generated by the energy and momentum that
are carried by the gravitational
field---Eq. \ref{eq:sch1}--- surrounding the point mass. This field
possesses an energy-momentum tensor\cite{wein}
\begin{eqnarray}
8\pi GT_{\mu\nu}^{\rm grav}&=&-{1\over 2}h^{(1)\lambda\kappa}\left[
\partial_\mu\partial_\nu
h^{(1)}_{\lambda\kappa}+\partial_\lambda\partial_\kappa h^{(1)}_{\mu\nu}
-\partial_\kappa\left(\partial_\nu
h^{(1)}_{\mu\lambda}+\partial_\mu
h^{(1)}_{\nu\lambda}\right)\right]\nonumber\\
&-&{1\over 2}\partial_\lambda h^{(1)}_{\sigma\nu}\partial^\lambda
h^{(1)\sigma}{}_\mu +{1\over 2}\partial_\lambda
h^{(1)}_{\sigma\nu}\partial^\sigma h^{(1)\lambda}{}_\mu -{1\over
4}\partial_\nu
h^{(1)}_{\sigma\lambda}\partial_\mu
h^{(1)\sigma\lambda}\nonumber\\
&-&{1\over 4}\eta_{\mu\nu}(\partial_\lambda h^{(1)}_{\sigma\chi}
\partial^\sigma h^{(1)\lambda\chi}
-{3\over 2}\partial_\lambda h^{(1)}_{\sigma\chi}\partial^\lambda
h^{(1)\sigma\chi})
-{1\over 4}h^{(1)}_{\mu\nu}\Box h^{(1)}\nonumber\\
&+&{1\over 2}\eta_{\mu\nu}h^{(1)\alpha\beta}\Box h^{(1)} _{\alpha\beta}
\end{eqnarray}
in terms of which the classical field correction to the point mass form
of the energy-momentum tensor is
determined to be
\begin{eqnarray}
T_{00}^{\rm grav}(r)&=&{1\over 8\pi G} \left(-{3\over
4}\vec{\nabla}f(r)\cdot\vec{\nabla}f(r)
-3f(r)\vec{\nabla}^2f(r)\right)+\ldots=-{3Gm^2\over 8\pi
r^4}+\ldots\nonumber\\
T_{ij}^{\rm grav}(r)&=&{1\over 8\pi G} \left(-{1\over
2}\nabla_if(r)\nabla_jf(r)+
{3\over
4}\delta_{ij}\vec{\nabla}f(r)\cdot\vec{\nabla}f(r)\right.\nonumber\\
&-&\left.f(r)\nabla_i\nabla_jf(r)+\delta_{ij}
f(r)\vec{\nabla}^2f(r)\right)+\ldots= -{7Gm^2\over 4\pi
r^4}\left({r_ir_j\over r^2}
-{1\over 2}\delta_{ij}\right)+\ldots\nonumber\\
\quad\label{eq:tao}
\end{eqnarray}
where the ellipses indicate contributions localized about the
origin. Obviously Eqs. \ref{eq:ta} and \ref{eq:tao} are identical,
demonstrating the correspondence of the nonanalytic $\sqrt{-q^2}$
terms and the classical field energy, just as found in I for the
electromagnetic case.

The remaining corrections to $T_{\mu\nu}$ contain an explicit factor of
$\hbar$ and are thus intrinsically quantum
mechanical in nature. The "physics'' behind these modifications can be
understood in terms of the position
uncertainty associated with quantum mechanics, which implies the
replacement of the distance $r$ in the classical
expression by the value $\sim r+{\hbar\over m}$.  Since for macroscopic
distances $\hbar/m<<r$, expansion of the
classical result in powers of $1/r$ leads qualitatively to the quantum
modifications found in our loop
calculation. We emphasize that both Eq. \ref{eq:ta},\ref{eq:tao} are long
range effects which arise only because
the graviton couples to a {\it massless} virtual particle---in this case
the self-interaction. The explicit factor
of $\hbar$ in the latter indicates clearly that these are quantum
effects whose strength and form are necessitated
by the quantum nature of the field theory.

\section{Classical terms in the metric}

Here we use this energy momentum tensor to calculate the associated
metric. In I we were able to show that this
procedure reproduced well-known results for classical metrics. We will
demonstrate the same feature for the
gravitational case. In this section we treat only the classical
$\sqrt{-q^2}$ terms, which we denote by $\sqrt{q}$
superscripts to the form factors. The method here is made somewhat more
complex by the necessity of dealing with
the nonlinearity of the Einstein equation. Here we must consistently
work to second order in $G$ and to this order
there is a nonlinear modification of the equations of motion relating
the energy momentum tensor and the metric.
This is worked out in the appendix---Eq. \ref{eq:eet}--- the result has
the form to second order in $G$
\begin{equation}
\Box h_{\mu\nu}^{(2)}=-16\pi G(T_{\mu\nu}^{\rm grav}-{1\over
2}\eta_{\mu\nu}
T^{\rm grav}) -\partial_\mu(f(r)\partial_\nu
f(r)) -\partial_\nu(f(r)\partial_\mu f(r)) \label{eq:sim}
\end{equation}
Noting that
\begin{eqnarray}
\nabla_i(f(r)\nabla_j f(r))&+&\nabla_j(f(r)\nabla_i f(r))=8G^2m^2\left(
4{r_ir_j\over r^6}-{\delta_{ij}\over r^4}\right)\nonumber\\
&=&4G^2m^2\nabla^2\left({\delta_{ij}\over r^2}-2{r_ir_j\over
r^4}\right)
\end{eqnarray}
we find then that
\begin{eqnarray}
h_{00}^{(2)}(r)&=&-16\pi G\int {d^3q\over
(2\pi)^3}e^{i\vec{q}\cdot\vec{r}} {1\over \vec{q}^2}\left({m\over
2}F_1^{\sqrt{q}}(-\vec{q}^2))-
{\vec{q}^2\over 4m}F_2^{\sqrt{q}}(-\vec{q}^2)\right)\nonumber\\
&=&-16\pi G\int {d^3q\over (2\pi)^3}e^{i\vec{q}\cdot\vec{r}}\left(
-{G\pi m^2\over 32 |\vec{q}|}-{7G\pi m^2\over 32|\vec{q}|}\right)
={2G^2m^2\over r^2}\nonumber\\
h_{0i}^{(2)}(r)&=&0\nonumber\\
h_{ij}^{(2)}(r)&=&-16\pi G\int {d^3q\over
(2\pi)^3}e^{i\vec{q}\cdot\vec{r}} {1\over \vec{q}^2}\left({m\over
2}F_1^{\sqrt{q}}(-\vec{q}^2)\delta_{ij} +{1\over 2m}(q_iq_j+{1\over
2}\delta_{ij}\vec{q}^2)
F_2^{\sqrt{q}}(-\vec{q}^2)\right)\nonumber\\
&+&4G^2m^2\left({\delta_{ij}\over r^2}-2{r_ir_j\over
r^4}\right)\nonumber\\
&=&-16\pi G\int {d^3q\over (2\pi)^3}e^{i\vec{q}\cdot\vec{r}}{1\over
\vec{q}^2} \left[\delta_{ij}\left(-{G\pi
m^2|\vec{q}|\over 32}+{7G\pi m^2|\vec{q}|\over 32}\right)
+{7G\pi m^2\over 16}{q_iq_j\over |\vec{q}|}\right]\nonumber\\
&+&4G^2m^2\left({\delta_{ij}\over r^2}-2{r_ir_j\over r^4}\right)=
-{G^2m^2\over r^2}\left(\delta_{ij}+{r_ir_j\over
r^2}\right)\label{eq:nai}
\end{eqnarray}
Comparing with the Schwarzschild solution in harmonic coordinates---Eq.
\ref{eq:sch}---we find complete agreement.

\section{Additional Quantum corrections to the metric}

Having identified the classical corrections, we could proceed in a
similar fashion to calculate the quantum
corrections using the $q^2\ln q^2$ non-analytic terms. However there is
one additional feature which needs to be
included. There is a quantum modification of the equations of motion,
which amounts to the addition of the vacuum
polarization diagram of Fig 2. In order to see that this is required,
let us look at the quantum corrected
effective action, which also has non-local long distance modifications.
At one loop one finds the effective action
\begin{equation}
Z[h] = -\int d^4x d^4y{1\over 2} \left[ h_{\mu\nu}(x) \right.
\left.\Delta^{\mu\nu,\alpha\beta}(x-y)h_{\alpha\beta}(y)+\textit{O}(h^3)\right]
+Z_{\rm matter}[h,\phi]
\end{equation}
Here the renormalized action $\Delta^{\mu\nu,\alpha\beta}(x-y)$ contains

\begin{equation}
\Delta^{\mu\nu,\alpha\beta}(x-y) = \delta^4(x-y)
D^{\mu\nu,\alpha\beta}_2 +\hat{\Pi}^{\mu\nu,\alpha\beta}(x-y)
+\textit{O}(\partial^4)
\end{equation}
where $D_2^{\mu\nu,\alpha\beta}$ is the differential operator
following from the Einstein action and
$\hat{\Pi}^{\mu\nu,\alpha\beta}(x-y)$ is the vacuum polarization
function after renormalization, see Fig. 1. Following the steps in
Appendix A we find that the vacuum polarization induces a change
in the equations of motion
\begin{eqnarray}
 \Box h_{\mu\nu}(x)&+& P_{\mu\nu,\alpha\beta}\int
d^4y\hat{\Pi}^{\alpha\beta,\gamma\delta}(x-y)h_{\gamma\delta}(y)\nonumber\\
 &=&-16\pi G(T^{\rm grav}_{\mu\nu} -{1\over 2}\eta_{\mu\nu} T^{\rm grav})
-\partial_\mu(f(r)\partial_\nu f(r))
-\partial_\nu(f(r)\partial_\mu f(r))\nonumber\\
\quad\label{eq:hmn}
\end{eqnarray}
where the projection operator $P_{\mu\nu,\alpha\beta}$ is given by
\begin{eqnarray}
P_{\mu\nu,\alpha\beta} &=& I_{\mu\nu,\alpha\beta} -{1\over 2}
\eta_{\mu\nu} \eta_{\alpha\beta} \nonumber \\
I_{\mu\nu,\alpha\beta} &=& {1\over 2}
(\eta_{\mu\alpha}\eta_{\nu\beta}+\eta_{\nu\alpha}\eta_{\mu\beta})
\end{eqnarray}

\begin{figure}
\begin{center}
\epsfig{file=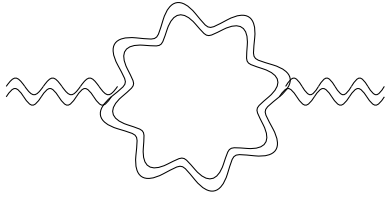,height=2.5cm,width=5cm}
\caption{The vacuum polarization diagram.}
\end{center}
\end{figure}

Eq. \ref{eq:hmn} can be written, in harmonic gauge, as
\begin{eqnarray}
\Box h_{\mu\nu} &=& -16\pi G(T_{\mu\nu}^{\rm grav}-{1\over 2}\eta_{\mu\nu}
T^{\rm grav})-\partial_\mu(f(r)
\partial_\nu f(r)) -\partial_\nu(f(r)\partial_\mu
f(r))\nonumber \\
&+&16\pi G\int d^4y
d^4z~P_{\mu\nu,\alpha\beta}\hat{\Pi}^{\alpha\beta,\gamma\delta}
(x-y)D(y-z)(T_{\gamma\delta}^{\rm matt}(z)-{1\over
2}\eta_{\gamma\delta} T^{\rm matt}(z))\nonumber\\
\quad
\end{eqnarray}
where the last term is just the vacuum polarization graph of Fig 2.

\begin{figure}
\begin{center}
\epsfig{file=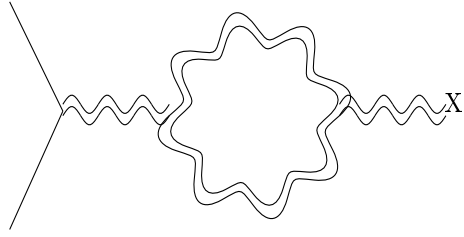,height=3cm,width=6cm}
\caption{Vacuum polarization modification of the energy-momentum
tensor.}
\end{center}
\end{figure}

The vacuum polarization has been calculated by 'tHooft and
Veltman\cite{thv}, and in momentum space it contains a
factor of $q^4\log(-q^2)$ which is the source of the nonlocality. The
specific form is
\begin{eqnarray}
\hat{\Pi}_{\alpha\beta,\gamma\delta}&=&-{2G\over
\pi}\log(-q^2)\left[{21\over 120}q^4
I_{\alpha\beta,\gamma\delta}+{23\over
120}q^4\eta_{\alpha\beta}\eta_{\gamma\delta} \right.\nonumber \\
&-&\left.{23\over 120}q^2(\eta_{\alpha\beta}q_\gamma q_\delta
+\eta_{\gamma\delta}q_\alpha
q_\beta)\right.\\
&-&\left.{21\over 240}q^2(q_\alpha q_\delta\eta_{\beta\gamma}+q_\beta
q_\delta\eta_{\alpha\gamma} +q_\alpha
q_\gamma\eta_{\beta\delta}+q_\beta q_\gamma\eta_{\alpha\delta})
+{11\over 30}q_\alpha q_\beta q_\gamma
q_\delta\right]\nonumber
\end{eqnarray}
When we employ this form along with the graviton propagator, we find for
that the vacuum polarization contributes
a shift in the metric
\begin{equation} \delta h^{(2){\rm vac pol}}_{\mu\nu}(x)=  32 G^2 \int
{d^3q
\over (2\pi)^3}
e^{i\vec{q}\cdot\vec{r}}\log (\vec{q}^2) \left[{21\over
120}T_{\mu\nu}^{\rm matt}(q) + \left({1\over 240}\eta_{\mu\nu}
-{11\over 60}{q^\mu q^\nu\over \vec{q}^2}\right)
   T^{\rm matt}(q)\right]
\end{equation}
In terms of components, we find,
\begin{eqnarray}
\delta h_{00}^{(2){\rm vac pol}} &=& - { 43G^2m\hbar  \over 15\pi r^3}
\nonumber \\
\delta h_{ij}^{(2){\rm vac pol}} &=& {G^2 m \hbar\over 15\pi r^3}
(\delta_{ij} +44 {r_i r_j\over r^2})
\end{eqnarray}

The remaining corrections come from the logarithms in the vertex
correction. Using the energy momentum tensor
shown above plus the integrals listed in the appendix we find
\begin{eqnarray}
\delta h_{00}^{(2){\rm vertex}} (r)&=&-16\pi G\int{d^3q\over
(2\pi)^3}e^{i\vec{q}\cdot\vec{r}} {1\over
\vec{q}^2}\left({m\over 2}F_1(-\vec{q}^2)
-{\vec{q}^2\over 4m}F_2(-\vec{q}^2)\right)\nonumber\\
&=&-16\pi G\int{d^3q\over (2\pi)^3}e^{i\vec{q}\cdot\vec{r}} {Gm\over
\pi}({3\over 8}+{1\over
2})\log\vec{q}^2={7G^2m\hbar\over \pi r^3}
\nonumber\\
\delta h_{0i}^{(2){\rm vertex}}(r)&=&0\nonumber\\
\delta h_{ij}^{(2){\rm vertex}}(r)&=&-16\pi G\int{d^3q\over
(2\pi)^3}e^{i\vec{q}\cdot\vec{r}} {1\over
\vec{q}^2}\log\vec{q}^2\left({m\over 2}
F_1(-\vec{q}^2))\delta_{ij}\right.\nonumber\\
&+&\left.{1\over 2m}(q_iq_j+{1\over 2}\delta_{ij}\vec{q}^2)
F_2(-\vec{q}^2)
\right)\nonumber\\
&=&-16\pi G\int{d^3q\over (2\pi)^3}e^{i\vec{q}\cdot\vec{r}}
{Gm\over \pi}\left(\delta_{ij}({3\over 8}-{1\over
2})-{q_iq_j\over \vec{q}^2}\right)\nonumber\\
&=&-{G^2m\hbar\over \pi r^3}\left(\delta_{ij}+8{r_ir_j\over r^2} \right)
\end{eqnarray}
where we have shown only the effects of the quantum logarithms. Adding
these corrections to the vacuum
polarization and classical terms reproduces the metric displayed in Eq.
3.

\section{Fermions and spin}

Having understood the spinless sector, we now turn our attention to the
case of a particle with spin, in
particular spin one-half.
 The general form for the spin 1/2 matrix element of the
   energy-momentum tensor can be written as\cite{pag}
   \begin{eqnarray}
   <p_2|T_{\mu\nu}|p_1>&=&\bar{u}(p_2)\left[ F_1(q^2)P_\mu P_\nu{1\over
   m}\right. \nonumber\\
   &-&F_2(q^2)({i\over 4m}\sigma_{\mu\lambda}q^\lambda
   P_\nu+{i\over 4m}\sigma_{\nu\lambda} q^\lambda P_\mu)\nonumber\\
   &+& \left. F_3(q^2)(q_\mu q_\nu-\eta_{\mu\nu}q^2){1\over m}\right]
u(p_1)
   \end{eqnarray}
The normalization condition $F_1(q^2=0)=1$ corresponds to
energy-momentum conservation as found before, while the
second normalization condition $F_2(q^2=0)=1$ is required by the
constraint of angular momentum conservation.
This can be seen by defining
\begin{eqnarray}
\hat{M}_{12}&=&\int d^3x(T_{01}x_2-T_{02}x_1)\nonumber\\
&&\stackrel{q\rightarrow 0}{\longrightarrow}-i(\nabla_{q})_2\int d^3x
e^{i\vec{q}\cdot\vec{r}}T_{01}(\vec{r})
+i(\nabla_{q})_1\int d^3xe^{i\vec{q}\cdot\vec{r}}T_{02}(\vec{r})
\end{eqnarray}
whereby
\begin{equation}
\lim_{q\rightarrow 0}<p_2|\hat{M}_{12}|p_1>={1\over 2}={1\over 2}
\bar{u}_\uparrow(p)\sigma_3u_\uparrow(p)F_2(q^2)
\end{equation}
{\it i.e.}, $F_2(q^2=0)=1$, as found explicitly in our calculation.

  The Feynman diagrams for fermions are shown in Fig 2.  We
find, as shown in the Appendix
   \begin{eqnarray}
   F_1(q^2)&=&1+{Gq^2\over \pi}\left({\pi^2m\over 16\sqrt{-q^2}}-
   {3\over 4}\log{-q^2\over m^2}\right) +\ldots \nonumber\\
   F_2(q^2)&=&1+{Gq^2\over \pi}\left({\pi^2m\over 4\sqrt{-q^2}}+
   {1\over 4}\log{-q^2\over m^2}\right) +\ldots\nonumber\\
   F_3(q^2)&=&{Gm^2\over \pi}\left({7\pi^2m\over
   16\sqrt{-q^2}}-\log{-q^2\over m^2}\right) +\ldots\label{eq:ff2}
   \end{eqnarray}

We convert this into an energy-momentum tensor.  Writing $\vec{S}=
\vec{\sigma}/2$ for the spin, the general
relation to the fermion form factors is
\begin{eqnarray}
    T_{00}(\vec{r})&=&\int{d^3q\over
   (2\pi)^3}e^{i\vec{q}\cdot\vec{r}}\left(mF_1(-\vec{q}^2)+
   {\vec{q}^2\over m}F_3(-\vec{q}^2)\right)\nonumber\\
    T_{0i}(\vec{r})&=&i\int{d^3q\over
   (2\pi)^3}e^{i\vec{q}\cdot\vec{r}}{1\over
   2}(\vec{S}\times\vec{q})_iF_2(-\vec{q}^2)\nonumber\\
   T_{ij}(\vec{r})&=&{1\over m}\int{d^3q\over
(2\pi)^3}e^{i\vec{q}\cdot\vec{r}}
(q_iq_j-\delta_{ij}\vec{q}^2)F_3(-\vec{q}^2)\label{eq:tc}
   \end{eqnarray}

Using our results Eq. \ref {eq:ff2} for the form factors this becomes
\begin{eqnarray}
   T_{00}(\vec{r})&=&\int{d^3q\over
   (2\pi)^3}e^{i\vec{q}\cdot\vec{r}}\left( m +{3Gm^2\pi \over
8}|\vec{q}|
  -{Gm\over 4\pi }
\vec{q}^2\log \vec{q}^2\right)+\ldots \nonumber \\
&=&m\delta^3(\vec{r}) -{3Gm^2\over 8\pi r^4}
-{3Gm\hbar\over 4\pi r^5}+\ldots\nonumber\\
    T_{0i}(\vec{r})&=&{i\over 2}\int{d^3q\over
(2\pi)^3}e^{i\vec{q}\cdot\vec{r}}(\vec{S}\times\vec{q})_i
\left(1-{Gm\pi\over
   4}|\vec{q}| - {G\over 4\pi}\vec{q}^2\log \vec{q}^2 \right)
+\ldots\nonumber \\
&=& {1\over 2}(\vec{S}\times\vec{\nabla})_i \delta^3(\vec{r})
+\left(-{Gm\over 2\pi r^6} +{15G\hbar\over
4\pi^2r^7} \right)(\vec{S}\times\vec{r})_i+\ldots \nonumber\\
   T_{ij}(\vec{r})&=&\int{d^3q\over
   (2\pi)^3}e^{i\vec{q}\cdot\vec{r}}\left(
{7Gm^2\pi\over
   16|\vec{q}|}-{Gm\over \pi}\log \vec{q}^2 \right)
\left(q_iq_j-\delta_{ij}\vec{q}^2\right)+\ldots
\nonumber \\
&=&-{7Gm^2\over 4\pi r^4}\left({r_ir_j\over r^2}-{1\over
   2}\delta_{ij}\right) +{2Gm\hbar\over
\pi^2r^5}\delta_{ij}+\ldots
   \quad\label{eq:tb}
\end{eqnarray}

We can again check the classical piece of this result against our
expectations of the energy-momentum carried by
the gravitational field. The spin-independent pieces are identical to
that found for the spinless case. In the
case of the off-diagonal component of the energy-momentum tensor, Eq.
\ref{eq:eet} yields
\begin{eqnarray}
T_{0i}^{\rm grav}&=&{1\over 8\pi G}\left(-{1\over 2}h^{(1)}_{0j}\nabla_i
\nabla_jh^{(1)}_{00}+{1\over
2}\nabla_{j}h^{(1)}_{ki}\nabla_kh^{(1)}_{j0} \right)
\nonumber\\
&=&{1\over 16\pi Gm}\left[-\left((\vec{S}\times
\vec{\nabla})_jf(r)\right)
\nabla_i\nabla_jf(r)+\left(\nabla_jf(r)\right)\nabla_i (\vec{S}\times
\vec{\nabla})_jf(r)\right]
\nonumber\\
&=&-{Gm\over 2\pi r^6}(\vec{S}\times\vec{r})_i
\end{eqnarray}
in agreement with the result obtained from Eq. \ref{eq:tb}

Now let us calculate the metric components.  In this case we find the
relation of the metric to the fermion form
factors is given by
 \begin{eqnarray}
    h_{00}(\vec{r})&=&-16\pi G\int{d^3q\over
   (2\pi)^3}e^{i\vec{q}\cdot\vec{r}}{1\over \vec{q}^2}\left( {m\over
2}F_1(-\vec{q}^2) -
{\vec{q}^2 \over 2m}F_3(-\vec{q}^2) \right)\nonumber\\
   h_{0i}(\vec{r})&=& -16\pi G{i\over 2}\int{d^3q\over
   (2\pi)^3}e^{i\vec{q}\cdot\vec{r}}{1\over \vec{q}^2}F_2(-\vec{q}^2)
   (\vec{S}\times\vec{q})_i
\nonumber\\
      h_{ij}(\vec{r})&=&-16\pi G\int{d^3q\over
   (2\pi)^3}e^{i\vec{q}\cdot\vec{r}} {1\over \vec{q}^2}\left(
{m\over 2}F_1(-\vec{q}^2)\delta_{ij} + {1\over m}(q_iq_j + {1\over 2}
\delta_{ij}
\vec{q}^2) F_3(-\vec{q}^2) \right)\nonumber\\
&+&{4G^2m^2\over r^2}\left(\delta_{ij}-2{r_ir_j\over
r^2}\right)\nonumber\\
\quad
   \end{eqnarray}
With the form factors calculated above this yields
\begin{eqnarray}
    h_{00}^{\rm vertex}(\vec{r})&=&-16\pi G\int{d^3q\over
   (2\pi)^3}e^{i\vec{q}\cdot\vec{r}}{1\over \vec{q}^2}\left( {m\over 2}
   -{Gm^2\pi|\vec{q}|\over 4}+{7Gm \vec{q}^2\over 8}\log \vec{q}^2
\right)+\ldots  \nonumber \\
   &=& -{2Gm \over r} +{2G^2m^2\over r^2}+{7G^2m\hbar\over  \pi r^3}
+\ldots\nonumber\\
   h_{0i}^{\rm vertex}(\vec{r})&=& -16\pi G{i\over 2}\int{d^3q\over
   (2\pi)^3}e^{i\vec{q}\cdot\vec{r}}{1\over \vec{q}^2}\left(1
 -{Gm\pi |\vec{q}|\over
   4} -{G \vec{q}^2 \over 4\pi}\log
\vec{q}^2\right)(\vec{S}\times\vec{q})_i
+\ldots\nonumber \\
&=& \left( {2G\over r^3} -{2G^2m\over r^4} + {3G^2\hbar\over \pi
r^5}\right)
(\vec{S}\times\vec{r})_i+\ldots\nonumber\\
   h_{ij}^{\rm vertex}(\vec{r})&=&-16\pi G\int{d^3q\over
   (2\pi)^3}e^{i\vec{q}\cdot\vec{r}}{1\over \vec{q}^2}\left( {m\over 2}
\delta_{ij} - ({Gm^2\pi|\vec{q}|\over 32}-{3Gm\vec{q}^2\over
8\pi}\log\vec{q}^2)\delta_{ij}
\right.\nonumber\\
&+&\left.(q_iq_j+{1\over 2}\vec{q}^2\delta_{ij})({7Gm^2\pi\over
16|\vec{q}|}-
{Gm\over \pi}\log{\vec{q}^2\over
m^2})\right)+{4G^2m^2\over r^2}\left(
\delta_{ij}-2{r_ir_j\over r^2}\right)+\ldots\nonumber \\
&=& -\delta_{ij}{2Gm\over r}-{G^2m^2\over r^2}\left(\delta_{ij}+
{r_ir_j\over r^2}\right)-{G^2 m\hbar\over \pi
r^3}\left(\delta_{ij} +8{r_ir_j\over r^2}\right)+\ldots\label{eq:rc1}
   \end{eqnarray}
We observe that the diagonal components of the vertex correction are
identical to those found
   for the spinless case, as expected, and that there exists a
   nonvanishing non-diagonal term associated with the spin.
The diagonal components of the vacuum polarization are also clearly
identical to the bosonic case, but there is a
new off-diagonal component associated with the spin
\begin{eqnarray}
h_{0i}^{(2){\rm vac pol}} &=&  32 G^2\int{d^3q\over
   (2\pi)^3}e^{i\vec{q}\cdot\vec{r}} \log \vec{q}^2 ~{21\over 240}
iF_2(q^2)(\vec{S}\times\vec{q})_i \nonumber \\
&=& {21G^2\hbar \over 5\pi r^5} (\vec{S}\times\vec{r})_i
\end{eqnarray}
Again these are added together in order to yield the result quoted in
the introduction. We have thus reproduced
the Kerr metric---Eq. \ref{eq:ker}---in harmonic gauge together with the
associated quantum corrections.

\section{Discussion of the metric and gravitational potential}

The quantum correction to the Schwarzschild metric has previously been
discussed by Duff\cite{duff}. While that
discussion properly identifies $\ln{-q^2}$ terms as the source of the
quantum effects, the calculation is
incomplete because it only includes the effect of the vacuum
polarization diagram. This can be traced to the
assumption of a  ``classical source'', which meant that the vertex
diagrams were not included. However, any source
has a gravitational field surrounding it and that field has a quantum
component. The effective field theory
treatment demonstrates the existence of quantum corrections due to the
vertex diagrams - they are of the same
order as those due to vacuum polarization and they must be included. In
this sense, there is no fully classical
source in gravity. If one takes the mass of a particle to infinity, the
gravitational coupling also grows and the
quantum effects do not decouple. Rather for a heavy mass it is long
distances which determines the classical
limit, as the quantum effects become smaller than the classical effects
in the limit of large distance. However,
the vertex corrections are as important as the vacuum polarization for
the quantum correction to the metric and
they must be included.

The bosonic diagrams that we have considered have also been parts of the
calculations of the quantum corrections
to the Newtonian potential. We have shown them in detail because there
has been numerical disagreements in the
literature. We believe that our results are the correct ones. There
appears to have been a numerical error in the
original result of Ref \cite{don}. We have identified the location of
that error and carefully reconsidered that
value. The identity of Eq. \ref{eq:bbid} makes it easy to repeat this
part of the calculation. The authors of Ref
\cite{akh} also appear to be in error. Their calculation would lead to
the wrong classical terms, which certainly
indicates an error and implies that the quoted quantum portion is also
not trustworthy. In addition, our fermionic
calculation serves as an independent confirmation of the bosonic result,
as the calculational details are quite
different even though the result is the same.

\begin{figure}
\begin{center}
\epsfig{file=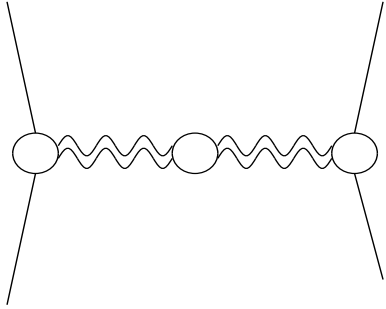,height=4cm,width=5cm}
\caption{Diagrams contributing to the one-particle-reducible potential.}
\end{center}
\end{figure}

If we use our present result to define the one-particle-reducible
potential, including the diagrams in Figure 3, we obtain the
result,
\begin{equation}
V({r})=-{Gm_1m_2\over r}\left(1-{G(m_1+m_2)\over {r}}-{167\over
30\pi}{G\hbar\over {r}^2}+\ldots\right)
\end{equation}
This potential is not itself the scattering potential. In a
separate work~\cite{nejbb} we calculate the other diagrams which
are required to fully define the scattering amplitude. These
include box diagrams and several triangle diagrams. However, the
1PR potential represents the sets of diagrams that are used to
define the running charge in QED and QCD and these diagrams can be
used for a similar definition here. We propose that the quantum
correction from these diagrams be used to define a running
gravitational coupling appropriate for harmonic gauge. This
results in
\begin{equation}
G(r) = G\left(1-{167\over 30\pi}{G\hbar\over {r}^2}\right)
\end{equation}
The fact that this definition is independent of the masses of the
objects involved suggests that it has a universal character
appropriate for the running charge. Our work shows that this form
is independent of spin. Note also that the charge becomes weaker
at shorter distances. This is in accord with a heuristic
expectation that the gravitational interaction at large distances
feels the total mass of the object, but when probed at small
distances gravity will see a smaller effect because the quantum
fluctuations spread out the energy contained in the fields. That
the running gravitational coupling varies with a power of $r$
rather than the logarithm is required by the dimensional
gravitational coupling constant.

\begin{center}

{\bf Acknowledgement}
\end{center}

This work was supported in part by the National Science Foundation under
award PHY-98-01875.  BRH would like to
acknowledge the warm hospitality of Forschungszentrum J\"{u}lich and
support by the Alexander von Humboldt
Foundation. N.E.J. Bjerrum-Bohr would like to thank P.H. Damgaard for 
discussions.

\section*{Appendix A - the equations of motion}
   While many details of the method are given in previous works, we do
make special use of the equations of motion
   in a way that is not shown elsewhere. Here we describe the background
that we need for our calculation.

 The full gravitational action is given by
\begin{equation}
S_g=\int d^4x \sqrt{-g}\left({1\over 16\pi G}R+{\cal
L}_m\right)\label{eq:ful}
\end{equation}
where ${\cal L}_m$ is the Lagrange density for matter.  Variation of Eq.
\ref{eq:ful} yields the Einstein equation
\begin{equation}
R_{\mu\nu}-{1\over 2}g_{\mu\nu}R=-8\pi GT_{\mu\nu}
\end{equation}
where the energy-momentum tensor $T_{\mu\nu}$ is given by
\begin{equation}
T_{\mu\nu}={2\over \sqrt{-g}}{\partial\over \partial
g^{\mu\nu}}(\sqrt{-g}{\cal L}_m)
\end{equation}
We work in the weak field limit, with an expansion in powers of the
gravitational coupling $G$
\begin{eqnarray}
g_{\mu\nu}&\equiv&\eta_{\mu\nu}+
h_{\mu\nu}^{(1)}+h_{\mu\nu}^{(2)}+\ldots\nonumber\\
g^{\mu\nu}&=&\eta^{\mu\nu}-h^{(1)\mu\nu}-h^{(2)\mu\nu}
+h^{(1)\mu\lambda} {h^{(1)}}_\lambda{}^\nu+\ldots
\end{eqnarray}
where here the superscript indicates the number of powers of $G$ which
appear and indices are understood to be
raised or lowered by $\eta_{\mu\nu}$. We shall also need the determinant
which is given by
\begin{equation}
\sqrt{-g}=\exp{1\over 2}{\rm tr}\log\,g=1+{1\over 2}(h^{(1)}+h^{(2)})
-{1\over
4}h^{(1)}_{\alpha\beta}h^{(1)\alpha\beta}+{1\over 8}h^{(1)2}+\ldots
\end{equation}
The corresponding curvatures are given by
\begin{eqnarray}
R_{\mu\nu}^{(1)}&=&{1\over 2}\left[\partial_\mu\partial_\nu h^{(1)}+
\partial_\lambda\partial^\lambda
h_{\mu\nu}^{(1)}-\partial_\mu\partial_\lambda {h^{(1)}}^\lambda{}_\nu
-\partial_\nu\partial_\lambda {h^{(1)}}^\lambda{}_\mu\right]\nonumber\\
R^{(1)}&=&\Box h^{(1)}-\partial_\mu\partial_\nu
h^{(1)\mu\nu}\nonumber\\
R_{\mu\nu}^{(2)}&=&{1\over 2}\left[\partial_\mu\partial_\nu h^{(2)}+
\partial_\lambda\partial^\lambda
h_{\mu\nu}^{(2)}-\partial_\mu\partial^\lambda h^{(2)}_{\lambda\nu}
-\partial_\nu\partial^\lambda h^{(2)}_{\lambda\mu}\right]\nonumber\\
&-&{1\over 4}\partial_\mu h^{(1)}_{\alpha\beta}\partial_\nu
h^{(1)\alpha\beta} -{1\over 2}\partial_\alpha
h^{(1)}_{\mu\lambda}\partial^\alpha h^{(1)}_{\lambda\nu}+{1\over
2}\partial_\alpha h^{(1)}_{\mu\lambda}
\partial^\lambda h^{(1)\alpha}_\nu\nonumber\\
&+&{1\over 2}h^{(1)\lambda\alpha}\left[\partial_\lambda\partial_\nu
h^{(1)}_{\mu\alpha}+\partial_\lambda\partial_\mu h^{(1)}_{\nu\alpha}-
\partial_\mu\partial_\nu h^{(1)}_{\lambda\alpha}-\partial_\lambda
\partial_\alpha h^{(1)}_{\mu\nu}\right]\nonumber\\
&+&{1\over 2}\left(\partial_\beta h^{(1)\beta\alpha}-{1\over 2}
\partial^\alpha h^{(1)}\right)\left(
\partial_\mu h^{(1)}_{\nu\alpha}+\partial_\nu h^{(1)}_{\mu\alpha}
-\partial_\alpha h^{(1)}_{\mu\nu}\right)\nonumber\\
R^{(2)}&=&\Box h^{(2)} -\partial^\mu\partial^\nu
h^{(2)}_{\mu\nu}-{3\over 4}\partial_\mu
h^{(1)}_{\alpha\beta}\partial^\mu h^{(1)\alpha\beta}+{1\over
2}\partial_\alpha h^{(1)}_{\mu\lambda}
\partial^\lambda h^{(1)\mu\alpha}\nonumber\\
&+&{1\over 2}h^{(1)\lambda\alpha}\left(2\partial_\lambda\partial^\beta
h^{(1)}_{\alpha\beta}-\Box
h^{(1)}_{\lambda\alpha}-
\partial_\lambda\partial_\alpha h^{(1)}\right)\nonumber\\
&+&\left(\partial^\beta h^{(1)\alpha}_\beta-{1\over 2}
\partial^\alpha h^{(1)}\right)
\left(\partial^\sigma h^{(1)}_{\sigma\alpha}-{1\over 2}
\partial_\alpha h^{(1)}\right)
\end{eqnarray}
In order to define the propagator, we must make a gauge choice and we
shall work in harmonic gauge---$g^{\mu\nu}
\Gamma^\lambda_{\mu\nu}=0$---which reads, to second order in the field
expansion,
\begin{eqnarray}
0&=&\partial^\beta h^{(1)}_{\beta\alpha}-{1\over 2}\partial_\alpha
h^{(1)}\nonumber\\
&=&\left(\partial^\beta h^{(2)}_{\beta\alpha}-{1\over 2}\partial_\alpha
h^{(2)}-{1\over 2}h^{(1)\lambda\sigma}
\left(\partial_\lambda h^{(1)}_{\sigma\alpha}+
\partial_\sigma h^{(1)}_{\lambda\alpha} -\partial_\alpha
h^{(1)}_{\sigma\lambda}\right)\right) \label{eq:gau}
\end{eqnarray}
Using these results, the Einstein equation reads, in lowest order,
\begin{equation}
\Box h^{(1)}_{\mu\nu}-{1\over 2}\eta_{\mu\nu}\Box h^{(1)}
-\partial_\mu\left(\partial^\beta h^{(1)}_{\beta\nu}-{1\over
2}\partial_\nu h^{(1)}\right)-\partial_\nu\left(\partial^\beta
h^{(1)}_{\beta\mu}-{1\over 2}\partial_\mu
h^{(1)}\right)=-16\pi GT_{\mu\nu}^{\rm matt}
\end{equation}
which, using the gauge condition Eq. \ref{eq:gau}a, can be written as
\begin{equation}
\Box \left(h_{\mu\nu}^{(1)}-{1\over 2}\eta_{\mu\nu}h^{(1)}\right)
=-16\pi G T_{\mu\nu}^{\rm matt}
\end{equation}
or in the equivalent form
\begin{equation}
\Box h_{\mu\nu}^{(1)}=-16\pi G\left( T_{\mu\nu}^{\rm matt}-{1\over
2}\eta_{\mu\nu} T^{\rm matt}\right)
\end{equation}
As shown in section 2, this equation has the familiar solution for a
static point mass
\begin{equation}
h_{\mu\nu}^{(1)}=\delta_{\mu\nu}f(r)\label{eq:los}
\end{equation}
where
$$f(r)=-{2Gm\over r}.$$

In second order the validity of the Einstein equation requires that
\begin{eqnarray}
R^{(2)}_{\mu\nu}-{1\over 2}\eta_{\mu\nu}R^{(2)}- {1\over
2}h^{(1)}_{\mu\nu}R^{(1)}=0
\end{eqnarray}
It is useful to write this equation in the form
\begin{equation}
\Box h^{(2)}_{\mu\nu}-{1\over 2}\eta_{\mu\nu}\Box h^{(2)}
-\partial_\mu\left(\partial^\beta h^{(2)}_{\beta\nu}-{1\over
2}\partial_\nu h^{(2)}\right)-\partial_\nu\left(\partial^\beta
h^{(2)}_{\beta\mu}-{1\over 2}\partial_\mu
h^{(2)}\right)\equiv-16\pi GT_{\mu\nu}^{\rm grav}\label{eq:soe}
\end{equation}
where $T^{\rm grav}_{\mu\nu}$ can be identified as the energy-momentum
carried by the gravitational field and can be
read off as
\begin{eqnarray}
8\pi GT_{\mu\nu}^{\rm grav}&=&-{1\over 2}h^{(1)\lambda\kappa}\left[
\partial_\mu\partial_\nu
h^{(1)}_{\lambda\kappa}+\partial_\lambda\partial_\kappa h^{(1)}_{\mu\nu}
-\partial_\kappa\left(\partial_\nu
h^{(1)}_{\mu\lambda}+\partial_\mu
h^{(1)}_{\nu\lambda}\right)\right]\nonumber\\
&-&{1\over 2}\partial_\lambda h^{(1)}_{\sigma\nu}\partial^\lambda
h^{(1)\sigma}{}_\mu +{1\over 2}\partial_\lambda
h^{(1)}_{\sigma\nu}\partial^\sigma h^{(1)\lambda}{}_\mu -{1\over
4}\partial_\nu
h^{(1)}_{\sigma\lambda}\partial_\mu
h^{(1)\sigma\lambda}\nonumber\\
&-&{1\over 4}\eta_{\mu\nu}(\partial_\lambda h^{(1)}_{\sigma\chi}
\partial^\sigma h^{(1)\lambda\chi}
-{3\over 2}\partial_\lambda h^{(1)}_{\sigma\chi}\partial^\lambda
h^{(1)\sigma\chi})
-{1\over 4}h^{(1)}_{\mu\nu}\Box h^{(1)}\nonumber\\
&+&{1\over 2}\eta_{\mu\nu}h^{(1)\alpha\beta}\Box h^{(1)}
_{\alpha\beta}\label{eq:gmn}
\end{eqnarray}
Using the gauge condition Eq. \ref{eq:gau}b, Eq. \ref{eq:soe} becomes
\begin{eqnarray}
\Box \left(h_{\mu\nu}^{(2)}\right.&-&\left.{1\over
2}\eta_{\mu\nu}h^{(2)}\right)
=-16\pi G T_{\mu\nu}^{\rm grav}\nonumber\\
&+&\partial_\mu \left( h^{(1)\lambda\sigma}\left(\partial_\lambda
h^{(1)}_{\sigma\nu}-
{1\over 2}\partial_\nu h^{(1)}_{\lambda\sigma}\right)\right)\nonumber\\
&+&\partial_\nu\left(
 h^{(1)\lambda\sigma}\left(\partial_\lambda h^{(1)}_{\sigma\mu}-
{1\over 2}\partial_\mu
h^{(1)}_{\lambda\sigma}\right)\right)\nonumber\\
&-&\eta_{\mu\nu}\partial^\alpha\left(h^{(1)\lambda\sigma}\left(\partial_\lambda
h_{\alpha\sigma}^{(1)}-{1\over 2}\partial_\alpha
h_{\lambda\sigma}^{(1)}\right)\right)
\end{eqnarray}
and, using the lowest order solution Eq. \ref{eq:los} we find the form
\begin{equation}
\Box h_{\mu\nu}^{(2)} =-16\pi G \left(T_{\mu\nu}^{\rm grav}-{1\over
2}\eta_{\mu\nu}T^{\rm grav}\right)
-\partial_\mu\left(f(r)\partial_\nu f(r)\right)
-\partial_\nu\left(f(r)\partial_\mu f(r)\right)\label{eq:eet}
\end{equation}
For use in the spin 1/2 case we note that the corresponding off-diagonal
equation reads
\begin{eqnarray}
\Box h_{0i}^{(2)}&=&-16\pi GT_{0i}^{\rm grav}-\nabla_i(h^{(1)}_{0j}
\nabla_jh_{00}^{(1)})
+\nabla_i(h^{(1)}_{jk}\nabla_jh^{(1)}_{0k}) \label{eq:eeu}
\end{eqnarray}
However, using the lowest order solutions found above we easily verify
that
\begin{equation}
\nabla_i(h^{(1)}_{0j}\nabla_jh_{00}^{(1)})=\nabla_i(h^{(1)}_{jk}
\nabla_jh^{(1)}_{0k})=0
\end{equation}
Thus the off-diagonal Einstein equation in second order has the simple
form
\begin{equation}
\Box h_{0i}^{(2)} =-16\pi G T_{0i}^{\rm grav}
\end{equation}
while the general form in second order is seen to be given by Eq.
\ref{eq:eet}.

\section*{Appendix B - details of the bosonic and fermionic vertex
corrections}
\subsection{Spin zero}

Here we show the calculation of the nonanalytic terms in vertex
correction, following the method of \cite{don}.
Such pieces arise from the diagrams in Figure 4, wherein the external
graviton couples to the massless graviton
fields in the loop. We have found that a symmetric ordering of the
momentum is useful, using the following
integrals
\begin{eqnarray}
I&=&\int{d^dk\over (2\pi)^d}{1\over (k-{q\over 2})^2(k+{q\over 2})^2
((p-k+{q\over 2})^2-m^2)}\nonumber\\
&=&{i\over 32\pi^2m^2}\left[-L-S\right]
+\ldots\nonumber\\
I_\mu&=&\int{d^dk\over (2\pi)^d}{k_\mu\over (k-{q\over 2})^2(k+{q\over
2})^2
((p-k+{q\over 2})^2-m^2)}\nonumber\\
&=&{i\over 32\pi^2m^2}\left[P_\mu\left(1+{q^2\over 2m}
\right)L+{q^2\over 4m^2}S\right]+\ldots\nonumber\\
I_{\mu\nu}&=&\int{d^dk\over (2\pi)^d}{k_\mu k_\nu\over (k-{q\over
2})^2(k+{q\over 2})^2
((p-k+{q\over 2})^2-m^2)}\nonumber\\
&=&{i\over 32\pi^2m^2}\left[-P_\mu P_\nu{q^2\over 2m^2} (L+{1\over
4}S)-\left(q_\mu
q_\nu-{\eta_{\mu\nu}q^2}\right) ({1\over 4}L+{1\over 8}S)\right]
+\ldots\nonumber\\
I_{\mu\nu\alpha}&=&\int{d^dk\over (2\pi)^d}{k_\mu k_\nu k_\alpha\over
(k-{q\over 2})^2(k+{q\over 2})^2
((p-k+{q\over 2})^2-m^2)}\nonumber\\
&=&{i\over 32\pi^2m^2}\left[P_\mu P_\nu P_\alpha\left(
{q^2\over 6m^2}\right)L\right.\nonumber\\
&+&\left.\left((q_\mu q_\nu-\eta_{\mu\nu}q^2)P_\alpha +(q_\mu
q_\alpha-\eta_{\mu\alpha}q^2)P_\nu+(q_\nu q_\alpha
-\eta_{\nu\alpha}q^2)P_\mu\right){L\over 12}\right]
+\ldots\nonumber\\
\quad
\end{eqnarray}
where $S=\pi^2m/\sqrt{-q^2},\,\,L=\log(-q^2/m^2)$.
\begin{figure}
\begin{center}
\epsfig{file=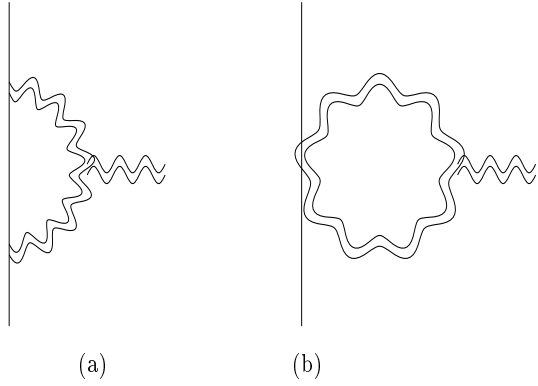,height=5cm,width=7cm} \caption{Gravitational
radiative correction diagrams leading to
nonanalytic components of form factors.}
\end{center}
\end{figure}
From Figure 4a, we have then
\begin{equation}
A^{\mu\nu}_{(a)}=iP^{\alpha,\lambda\kappa}iP^{\gamma\delta,\rho\sigma}
i\int {d^4\ell\over
(2\pi)^4}{\tau_{\alpha\beta}(p,p'-\ell) \tau_{\gamma\delta}(p'-\ell,p')
\tau^{\mu\nu}_{\rho\sigma,\lambda\kappa}(\ell,q)\over \ell^2(\ell-q)^2
((\ell-p')^2-m^2)}\label{eq:eqa}
\end{equation}
while from Figure 4b,
\begin{equation}
A^{\mu\nu}_{(b)}={i\over 2}
P^{\alpha\beta,\lambda\kappa}iP^{\gamma\delta,\rho\sigma}
\tau_{\alpha\beta,\gamma\delta}(p,p')\int{d^4\ell\over (2\pi)^4}
{\tau^{\mu\nu}_{\lambda\kappa,\rho\sigma}(\ell,q)\over
\ell^2(\ell-q)^2}\label{eq:eqb}
\end{equation}

Here the coupling to matter via one-graviton and two-graviton vertices
can be found by expanding the spin zero matter Lagrangian
\begin{equation}
\sqrt{-g}{\cal L}_m=\sqrt{-g}\left({1\over 2}D_\mu\phi g^{\mu\nu}D_\nu\phi
-{1\over 2}m^2\phi^2\right)
\end{equation}
via
\begin{eqnarray}
\sqrt{-g}{\cal L}_m^{(0)}&=&{1\over
2}(\partial_\mu\phi\partial^\mu\phi-m^2\phi^2)
\nonumber\\
\sqrt{-g}{\cal L}_m^{(1)}&=&-{1\over 2}h^{(1)\mu\nu}\left(\partial_\mu\phi
\partial_\nu\phi-{1\over 2}\eta_{\mu\nu}(\partial_\alpha\phi
\partial^\alpha\phi-m^2\phi^2)\right)\nonumber\\
\sqrt{-g}{\cal L}_m^{(2)}&=&-{1\over 2}h^{(2)\mu\nu}\left(\partial_\mu\phi
\partial_\nu\phi-{1\over 2}\eta_{\mu\nu}(\partial_\alpha\phi
\partial^\alpha\phi-m^2\phi^2)\right)\nonumber\\
&+&{1\over 2}\left(h^{(1)\mu\lambda}{h^{(1)\nu}}_\lambda
-{1\over
2}h^{(1)}h^{(1)\mu\nu}\right)\partial_\mu\phi\partial_\nu\phi\nonumber\\
&-&{1\over 8}\left(h^{(1)\alpha\beta}h^{(1)}_{\alpha\beta}-{1\over 2}
h^{(1)2}\right)(\partial^\alpha\phi\partial_\alpha\phi-m^2\phi^2)\label{eq:mlg}
\end{eqnarray}
The one- and two-graviton vertices are then respectively
\begin{eqnarray}
\tau_{\alpha\beta}(p,p')&=&{-i\kappa\over
2}\left(p_\alpha{p'}_\beta+{p'}_\alpha
p_\beta-\eta_{\alpha\beta}(p\cdot p'-m^2)\right)\nonumber\\
\tau_{\alpha\beta,\gamma\delta}(p,p')&=&i\kappa^2\left[
I_{\alpha\beta,\rho\xi}I^\xi{}_{\sigma,\gamma\delta}
\left(p^\rho{p'}^\sigma+{p'}^\rho p^\sigma\right)\right.\nonumber\\
&-&\left.{1\over
2}\left(\eta_{\alpha\beta}I_{\rho\sigma,\gamma\delta}
+\eta_{\gamma\delta}I_{\rho\sigma,\alpha\beta}\right){p'}^\rho
p^\sigma\right.\nonumber\\
&-&\left.{1\over 2}\left(I_{\alpha\beta,\gamma\delta} -{1\over
2}\eta_{\alpha\beta}\eta_{\gamma\delta}\right) \left(p\cdot
p'-m^2\right)\right]
\end{eqnarray}
where we have defined $\kappa^2=32\pi G$.
We also require the triple graviton vertex
$\tau_{\alpha\beta,\gamma\delta}^{\mu\nu}(k,q)$ whose form is
\begin{eqnarray}
\tau^{\mu\nu}_{\alpha\beta,\gamma\delta}(k,q)&=&{i\kappa\over 2}\left\{
P_{\alpha\beta,\gamma\delta}
\left[k^\mu k^\nu+(k-q)^\mu (k-q)^\nu+q^\mu q^\nu-{3\over
2}\eta^{\mu\nu}q^2\right]\right.\nonumber\\
&+&\left.2q_\lambda
q_\sigma\left[I^{\lambda\sigma,}{}_{\alpha\beta}I^{\mu\nu,}
{}_{\gamma\delta}+I^{\lambda\sigma,}{}_{\gamma\delta}I^{\mu\nu,}
{}_{\alpha\beta}-I^{\lambda\mu,}{}_{\alpha\beta}I^{\sigma\nu,}
{}_{\gamma\delta}-I^{\sigma\nu,}{}_{\alpha\beta}I^{\lambda\mu,}
{}_{\gamma\delta}\right]\right.\nonumber\\
&+&\left.[q_\lambda
q^\mu(\eta_{\alpha\beta}I^{\lambda\nu,}{}_{\gamma\delta}
+\eta_{\gamma\delta}I^{\lambda\nu,}{}_{\alpha\beta})+ q_\lambda
q^\nu(\eta_{\alpha\beta}I^{\lambda\mu,}{}_{\gamma\delta}
+\eta_{\gamma\delta}I^{\lambda\mu,}{}_{\alpha\beta})\right.\nonumber\\
&-&\left.q^2(\eta_{\alpha\beta}I^{\mu\nu,}{}_{\gamma\delta}+\eta_{\gamma\delta}
I^{\mu\nu,}{}_{\alpha\beta})-\eta^{\mu\nu}q^\lambda
q^\sigma(\eta_{\alpha\beta}
I_{\gamma\delta,\lambda\sigma}+\eta_{\gamma\delta}
I_{\alpha\beta,\lambda\sigma})]\right.\nonumber\\
&+&\left.[2q^\lambda(I^{\sigma\nu,}{}_{\alpha\beta}
I_{\gamma\delta,\lambda\sigma}(k-q)^\mu
+I^{\sigma\mu,}{}_{\alpha\beta}I_{\gamma\delta,\lambda\sigma}(k-q)^\nu\right.\nonumber\\
&-&\left.I^{\sigma\nu,}{}_{\gamma\delta}I_{\alpha\beta,\lambda\sigma}k^\mu-
I^{\sigma\mu,}{}_{\gamma\delta}I_{\alpha\beta,\lambda\sigma}k^\nu)\right.\nonumber\\
&+&\left.q^2(I^{\sigma\mu,}{}_{\alpha\beta}I_{\gamma\delta,\sigma}{}^\nu+
I_{\alpha\beta,\sigma}{}^\nu
I^{\sigma\mu,}{}_{\gamma\delta})+\eta^{\mu\nu}q^\lambda q_\sigma
(I_{\alpha\beta,\lambda\rho}I^{\rho\sigma,}{}_{\gamma\delta}+
I_{\gamma\delta,\lambda\rho}I^{\rho\sigma,}{}_{\alpha\beta})]\right.\nonumber\\
&+&\left.[(k^2+(k-q)^2)\left(I^{\sigma\mu,}{}_{\alpha\beta}I_{\gamma\delta,\sigma}{}^\nu
+I^{\sigma\nu,}{}_{\alpha\beta}I_{\gamma\delta,\sigma}{}^\mu-{1\over
2}\eta^{\mu\nu}P_{\alpha\beta,\gamma\delta}\right)\right.\nonumber\\
&-&\left.(k^2\eta_{\gamma\delta}I^{\mu\nu,}{}_{\alpha\beta}+(k-q)^2\eta_{\alpha\beta}
I^{\mu\nu,}{}_{\gamma\delta})]\right\}
\end{eqnarray}

Before presenting our results, we note a simplification---it can be
easily seen that the terms in the 3-graviton
vertex function proportional to $k^2$ or $(k-q)^2$ do {\it not} produce
nonanalytic pieces when inserted into
either Eq. \ref{eq:eqa} or Eq. \ref{eq:eqb} and can be dropped.

A further enormous simplification of indices results from the
identity\cite{bohr}
\begin{equation}
P^{\xi\zeta,\alpha\beta}{\tau^{\mu\nu}}_{\alpha\beta,\gamma\delta}(k,q)
P^{\gamma\delta,\kappa\rho}=\tau^{\mu\nu,\xi\zeta,\kappa\rho}(k,q)\label{eq:bbid}
\end{equation}
for all the terms which lead to nonanalytic corrections. This can
be verified straightforwardly.  The resulting integrals are still
tedious, but can be done directly.

 Decomposing the remaining piece of this vertex
into the four bracketed terms, we list our results in terms of the
contributions from each bracket separately:
\begin{eqnarray}
{\rm Figure}\,\,4a:\,\,\,F_1(q^2)&=&{Gq^2\over \pi}\left([{1\over
4}-2+1+0]\log(-q^2)
+[{1\over 16}-1+1+0]{\pi^2m\over \sqrt{-q^2}}\right)\nonumber\\
&=&{Gq^2\over \pi}\left(-{3\over 4}\log(-q^2)+{1\over 16}{\pi^2m\over
\sqrt{-q^2}}\right)\nonumber\\
F_2(q^2)&=&{Gm^2\over \pi}\left([{13\over 3}-1+0-1]\log(-q^2)
+[{7\over 8}-1+2-1]{\pi^2m\over \sqrt{-q^2}}\right)\nonumber\\
&=&{Gm^2\over \pi}\left({7\over 3}\log(-q^2)+{7\over 8}{\pi^2m^2\over
\sqrt{-q^2}}\right)\nonumber\\
{\rm Figure}\,\,4b:\,\,\,F_1(q^2)&=&{Gq^2\over \pi}\left(
[0+2+0-2]\log(-q^2)\right)\nonumber\\
&=&0\nonumber\\
F_2(q^2)&=&{Gm^2\over \pi}\left(
[-{25\over 3}+0+2+2]\log(-q^2)\right)\nonumber\\
&=&{Gm^2\over \pi}\left(-{13\over 3}\log(-q^2)\right)
\end{eqnarray}

\subsection{Spin 1/2}

For the case of spin 1/2 we require some additional formalism in order
to extract the gravitational couplings.  In
this case the matter Lagrangian reads
\begin{equation}
\sqrt{e}{\cal L}_m=\sqrt{e}\bar{\psi}(i\gamma^a{e_a}^\mu D_\mu-m)\psi
\end{equation}
and involves the vierbein ${e_a}^\mu$ which links global coordinates
with those in a locally flat space.  The
vierbein is in some sense the ``square root'' of the metric tensor
$g_{\mu\nu}$ and satisfies the relations
\begin{eqnarray}
{e^a}_\mu {e^b}_\nu\eta_{ab}&=&g_{\mu\nu}\quad{e^a}_\mu
e_{a\nu}=g_{\mu\nu}\nonumber\\
e^{a\mu}e_{b\mu}&=&\delta^a_b\quad e^{a\mu}{e_a}^\nu=g^{\mu\nu}
\end{eqnarray}
The covariant derivative is defined via
\begin{equation}
D_\mu\psi=\partial_\mu\psi+{i\over 4}\sigma^{ab}\omega_{\mu ab}
\end{equation}
where
\begin{eqnarray}
\omega_{\mu ab}&=&{1\over 2}{e_a}^\nu(\partial_\mu e_{b\nu}-\partial_\nu
e_{b\mu})-{1\over
2}{e_b}^\nu(\partial_\mu
e_{a\nu}-\partial_\nu e_{a\mu})\nonumber\\
&+&{1\over 2}{e_a}^\rho{e_b}^\sigma(\partial_\sigma
e_{c\rho}-\partial_\rho e_{c\sigma}){e_\mu}^c
\end{eqnarray}
The connection with the metric tensor can be made via the expansion
\begin{equation}
{e^a}_\mu=\delta^a_\mu+c^{(1)a}_\mu+c^{(2)a}_\mu+\ldots
\end{equation}
where, as before, the superscript indicates the number of powers of the
gravitational coupling $G$ which are
present.  The inverse of this matrix is
\begin{equation}
{e_a}^\mu=\delta_a^\mu-c_a^{(1)\mu}-c_a^{(2)\mu}+{c^{(1)\mu}_b{c^{(1)b}_a+\ldots}}
\end{equation}
and we find
\begin{eqnarray}
g_{\mu\nu}&=&\eta_{\mu\nu}+c^{(1)}_{\mu\nu}+c^{(1)}_{\nu\mu}+c^{(2)}_{\mu\nu}
+c^{(2)}_{\nu\mu}\nonumber\\
&+&{c^{(1)a}}_\mu c^{(1)}_{a\nu}+\ldots\nonumber\\
g^{\mu\nu}&=&\eta^{\mu\nu}-c^{(1)\mu\nu}-c^{(1)\nu\mu}-c^{(2)\mu\nu}
-c^{(2)\nu\mu}\nonumber\\
&+&c^{(1)a\mu} c^{(1)\nu}_a+c^{(1)\mu a} {c^{(1)}_a}^\nu+ c^{(1)\mu a}
c^{(1)\nu}_a+\ldots
\end{eqnarray}
For our purposes we shall use only the symmetric component of the
c-matrices, since these are physical and can be
connected to the metric tensor, while their antisymmetric components are
associated with freedom of homogeneous
transformations of the local Lorentz frames and do not contribute to
nonanalyticity.  We find then
$$c^{(1)}_{\mu\nu}\rightarrow {1\over
2}(c^{(1)}_{\mu\nu}+c^{(1)}_{\nu\mu})={1\over 2}h^{(1)}_{\mu\nu}$$ We
have then
\begin{eqnarray}
{\rm det}\,e&=&1+c+{1\over 2}c^2-{1\over
2}{c_a}^b{c_b}^a+\ldots\nonumber\\
&=&1+{1\over 2}h+{1\over 8}h^2-{1\over 8}{h_a}^b{h_b}^a+\ldots
\end{eqnarray}
Using these forms the matter Lagrangian has the expansion
\begin{eqnarray}
\sqrt{e}{\cal L}_m^{(0)}&=&\bar{\psi}({i\over
2}\gamma^\alpha\delta^\mu_\alpha\partial_\mu^{LR}-m)\psi\nonumber\\
\sqrt{e}{\cal L}_m^{(1)}&=&-{1\over
2}h^{(1)\alpha\beta}\bar{\psi}i\gamma_\alpha
\partial_\beta^{LR}\psi-{1\over 2}h^{(1)}\bar{\psi}({i\over
2}\not\!{\partial}^{LR}-m)\psi\nonumber\\
\sqrt{e}{\cal L}_m^{(2)}&=&-{1\over
2}h^{(2)\alpha\beta}\bar{\psi}i\gamma_\alpha
\partial_\beta^{LR}\psi-{1\over 2}h^{(2)}\bar{\psi}({i\over
2}\not\!{\partial}^{LR}-m)\psi\nonumber\\
&-&{1\over 8}h^{(1)}_{\alpha\beta}
h^{(1)\alpha\beta}\bar{\psi}i\gamma^\gamma\partial_\lambda^{LR}\psi+{1\over
16}(h^{(1)})^2\bar{\psi}i\gamma^\gamma\partial_\gamma^{LR}\psi\nonumber\\
&-&{1\over
8}h^{(1)}\bar{\psi}i\gamma^\alpha{h_\alpha}^\lambda\partial_\lambda^{LR}\psi
+{3\over
16}h_{\delta\alpha}^{(1)}h^{(1)\alpha\mu}\bar{\psi}i\gamma^\delta\partial_\mu^{LR}\psi\nonumber\\
&+&{1\over 4}h_{\alpha\beta}^{(1)}h^{(1)\alpha\beta}\bar{\psi}
m\psi-{1\over
8}(h^{(1)})^2\bar{\psi}m\psi\nonumber\\
&+&{i\over 16}h_{\delta\nu}^{(1)}(\partial_\beta
h^{(1)\nu}_\alpha-\partial_\alpha h^{(1)\nu}_\beta)
\epsilon^{\alpha\beta\delta\epsilon}\bar{\psi}\gamma_\epsilon\gamma_5\psi
\end{eqnarray}
where
$$\bar{\psi}\partial_\alpha^{LR}\psi\equiv \bar{\psi}\partial_\alpha\psi-
(\partial_\alpha\bar{\psi})\psi .$$
 The corresponding one- and two-graviton vertices are found
then to be
\begin{eqnarray}
\tau_{\alpha\beta}(p,p')&=&{-i\kappa\over 2}\left[{1\over
4}(\gamma_\alpha(p+p')_\beta+\gamma_\beta(p+p')_\alpha)-{1\over
2}\eta_{\alpha\beta}({1\over
2}(\not\!\!{p}+\not\!\!{p}')-m)\right]\nonumber\\
\tau_{\alpha\beta,\gamma\delta}(p,p')&=&i\kappa^2\left\{-{1\over
2}({1\over
2}(\not\!\!{p}+\not\!\!{p}')-m)P_{\alpha\beta,\gamma\delta}\right.\nonumber\\
&-&\left.{1\over
16}[\eta_{\alpha\beta}(\gamma_\gamma(p+p')_\delta+\gamma_\delta(p+p')_\gamma)
\right.\nonumber\\
&+&\left.\eta_{\gamma\delta}(\gamma_\alpha
(p+p')_\beta+\gamma_\beta(p+p')_\alpha)]\right.\nonumber\\
&+&\left.{3\over
16}(p+p')^{\epsilon}\gamma^{\xi}(I_{\xi\phi,\alpha\beta}{I^{\phi}}_{\epsilon,\gamma\delta}
+I_{\xi\phi,\gamma\delta}{I^{\phi}}_{\epsilon,\alpha\beta})\right.\nonumber\\
&+&\left.{i\over 8}\epsilon^{\rho\sigma\eta\lambda}\gamma_\lambda
\gamma_5({I_{\alpha\beta,\eta}}^\nu
I_{\gamma\delta,\sigma\nu}{k'}_\rho-{I_{\gamma\delta,\eta}}^\nu
I_{\alpha\beta,\sigma\nu}k_\rho)\right\}
\end{eqnarray}

With these results in hand the loop integrations can now be performed,
as before, yielding, for spin 1/2
\begin{eqnarray}
{\rm Figure}\,\,4a:\,\,\,F_1(q^2)&=&{Gq^2\over \pi} \left([{1\over
4}-{3\over 4}-{1\over 2}+{1\over 4}]\log(-q^2)
+[{1\over 16}+0-1+1]{\pi^2m\over \sqrt{-q^2}}\right)\nonumber\\
&=&{Gq^2\over \pi}\left(-{3\over 4}\log(-q^2)+{1\over 16}{\pi^2m\over
\sqrt{-q^2}}\right)\nonumber\\
F_2(q^2)&=&{Gq^2\over \pi} \left([{1\over 6}-{7\over 4}+{3\over
4}+{13\over 12}]\log(-q^2)
+[0+0-{1\over 2}+{3\over 4}]{\pi^2m\over \sqrt{-q^2}}\right)\nonumber\\
&=&{Gq^2\over \pi}\left({1\over 4}\log(-q^2)+{1\over 4}{\pi^2m\over
\sqrt{-q^2}}\right)\nonumber\\
F_3(q^2)&=&{Gm^2\over \pi}\left([{4\over 3}+0-1+{1\over 4}]\log(-q^2)
+[{7\over 16}+0+0+0]{\pi^2m\over \sqrt{-q^2}}\right)\nonumber\\
&=&{Gm^2\over \pi}\left({7\over 12}\log(-q^2)+{7\over 16}
{\pi^2m^2\over \sqrt{-q^2}}\right)\nonumber\\
{\rm Figure}\,\,4b:\,\,\,F_1(q^2)&=&{Gq^2\over \pi}\left(
[0+{11\over 4}-{1\over 2}-{9\over 4}]\log(-q^2)\right)\nonumber\\
&=&0\nonumber\\
F_2(q^2)&=&{Gq^2\over \pi}\left(
[0+{7\over 4}-{1\over 2}-{5\over 4}]\log(-q^2)\right)\nonumber\\
&=&0\nonumber\\
F_3(q^2)&=&{Gm^2\over \pi}\left(
[-{10\over 3}+0+0+{7\over 4}]\log(-q^2)\right)\nonumber\\
&=&-{Gm^2\over \pi}{19\over 12}\log(-q^2)
\end{eqnarray}
All calculations of the form factors where done by hand and by computer.
To do the various contractions of indices and integrations by
computer, a computer algorithm for Maple 7
(TM)\footnote{\footnotesize{Maple and Maple V are registered trademarks of
Waterloo Maple Inc.}} were developed and used to do the calculations.

\section*{Appendix C - useful integrals}
Here we collect the integrals used to calculate the long range
corrections to the energy momentum tensor and the
metric. For the classical correction to the energy momentum tensor we
use
\begin{eqnarray}
& &\int{d^3q\over (2\pi)^3}e^{i\vec{q}\cdot\vec{r}}|\vec{q}|=-{1\over
\pi^2r^4}\nonumber\\
&
&\int{d^3q\over(2\pi)^3}e^{i\vec{q}\cdot\vec{r}}q_j|\vec{q}|={-4ir_j\over
\pi^2r^6}\nonumber\\
& &\int{d^3q\over (2\pi)^3}e^{i\vec{q}\cdot\vec{r}}{q_iq_j\over
|\vec{q}|} = {1\over
\pi^2r^4}\left(\delta_{ij}-4{r_ir_j\over r^2}\right)\label{eq:r1}
\end{eqnarray}
and the quantum effect use
\begin{eqnarray}
& &\int{d^3q\over (2\pi)^3}e^{i\vec{q}\cdot\vec{r}}\vec{q}^2\log
\vec{q}^2 ={3\over \pi r^5}\nonumber\\
& &  \int{d^3q\over (2\pi)^3}e^{i\vec{q}\cdot\vec{r}}{q_j\vec{q}^2\log
\vec{q}^2}={i15r_j\over \pi r^7}\nonumber\\
& & \int{d^3q\over (2\pi)^3}e^{i\vec{q}\cdot\vec{r}}q_iq_j\log
\vec{q}^2={1\over \pi r^5}\delta_{ij}\label{eq:r2}
\end{eqnarray}
For the metric we require
\begin{eqnarray}
& &   \int{d^3q\over
   (2\pi)^3}e^{i\vec{q}\cdot\vec{r}}{1\over |\vec{q}|}
={1\over 2\pi^2r^2} \nonumber \\
 & &   \int{d^3q\over
   (2\pi)^3}e^{i\vec{q}\cdot\vec{r}}{q_j\over |\vec{q}|}
={ir_j\over \pi^2r^4}\nonumber\\
& &   \int{d^3q\over
   (2\pi)^3}e^{i\vec{q}\cdot\vec{r}}{q_iq_j\over |\vec{q}|^3}={1\over
   2\pi^2r^2}\left(\delta_{ij}-2{r_ir_j\over r^2}\right)\label{eq:r3}
   \end{eqnarray}
and
\begin{eqnarray}
& &\int{d^3q\over (2\pi)^3}e^{i\vec{q}\cdot\vec{r}}
\log\vec{q}^2=-{1\over 2\pi r^3} \nonumber \\
& &\int{d^3q\over (2\pi)^3}e^{i\vec{q}\cdot\vec{r}}
q_j\log\vec{q}^2={-i3r_j\over 2\pi
r^5}\nonumber\\
& &\int{d^3q\over (2\pi)^3}e^{i\vec{q}\cdot\vec{r}}
\left({q_iq_j\over
\vec{q}^2}\right)\log\vec{q}^2 =-{r_ir_j\over 2\pi r^5}
\end{eqnarray}

\end{document}